\documentclass[prl,aps,epsfig,superscriptaddress,twocolumn]{revtex4-1}
\setcitestyle{super}
\usepackage{amsfonts,amssymb,amscd,amsthm}
\usepackage{graphicx}
\usepackage{mathrsfs}
\usepackage[intlimits]{amsmath}
\usepackage[colorlinks, citecolor=red]{hyperref}
\usepackage{lineno}
\usepackage{etoolbox}
\begin{document}

\title{Realizing coherently convertible dual-type qubits with the same ion species}

\author{H.-X. Yang}
\thanks{These authors contribute equally to this work}
\affiliation{Center for Quantum Information, Institute for Interdisciplinary Information Sciences, Tsinghua University, Beijing 100084, PR China}

\author{J.-Y. Ma}
\thanks{These authors contribute equally to this work}
\affiliation{Center for Quantum Information, Institute for Interdisciplinary Information Sciences, Tsinghua University, Beijing 100084, PR China}

\author{Y.-K. Wu}
\affiliation{Center for Quantum Information, Institute for Interdisciplinary Information Sciences, Tsinghua University, Beijing 100084, PR China}

\author{Y. Wang}
\affiliation{Center for Quantum Information, Institute for Interdisciplinary Information Sciences, Tsinghua University, Beijing 100084, PR China}

\author{M.-M. Cao}
\affiliation{Center for Quantum Information, Institute for Interdisciplinary Information Sciences, Tsinghua University, Beijing 100084, PR China}

\author{W.-X. Guo}
\affiliation{Center for Quantum Information, Institute for Interdisciplinary Information Sciences, Tsinghua University, Beijing 100084, PR China}

\author{Y.-Y. Huang}
\affiliation{Center for Quantum Information, Institute for Interdisciplinary Information Sciences, Tsinghua University, Beijing 100084, PR China}

\author{L. Feng}
\affiliation{Center for Quantum Information, Institute for Interdisciplinary Information Sciences, Tsinghua University, Beijing 100084, PR China}

\author{Z.-C. Zhou}
\affiliation{Center for Quantum Information, Institute for Interdisciplinary Information Sciences, Tsinghua University, Beijing 100084, PR China}

\author{L.-M. Duan}
\email{Corresponding author: lmduan@tsinghua.edu.cn}
\affiliation{Center for Quantum Information, Institute for Interdisciplinary Information Sciences, Tsinghua University, Beijing 100084, PR China}


\maketitle

\textbf{Trapped ions constitute one of the most promising systems for implementing quantum computing and networking \cite{nielsen2000quantum, Monroe2013}.
For large-scale ion-trap-based quantum computers and networks, it is critical to have two types of qubits, one for computation and storage, while the other for auxiliary operations like runtime qubit detection \cite{negnevitsky2018repeated}, sympathetic cooling \cite{Rohde_2001,PhysRevA.65.040304,PhysRevA.68.042302,PhysRevA.79.050305}, and repetitive entanglement generation through photon links \cite{duan2010colloquium,hucul2015modular}. Dual-type qubits have previously been realized in hybrid systems using two ion species \cite{tan2015multi,inlek2017multispecies,negnevitsky2018repeated,bruzewicz2019dual,Wan875}, which, however, introduces significant experimental challenges for laser setup, gate operations \cite{sosnova2021character} as well as the control of the fraction and positioning of each qubit type within an ion crystal \cite{lin2016sympathetic}. Here we solve these problems by implementing two coherently-convertible qubit types using the same ion species. We encode the qubits into two pairs of clock states of the ${}^{171}\mathrm{Yb}^+$ ions, and achieve fast and high-fidelity conversion between the two types using narrow-band lasers. We further demonstrate that operations on one qubit type, including sympathetic laser cooling, gates and qubit detection, have crosstalk errors less than $0.03\%$ on the other type, well below the error threshold for fault-tolerant quantum computing. Our work showcases the feasibility and advantages of using coherently convertible dual-type qubits with the same ion species for future large-scale quantum computing and networking.}

Quantum computers have attracted wide research interest owing to the potential exponential speedup over any classical computers on certain tasks like factorizing large integers and quantum simulation of material properties \cite{nielsen2000quantum}. However, quantum states are also fragile and requires quantum error correction to protect against environmental noise and control errors \cite{nielsen2000quantum,campbell2017roads}. It is thus crucial to have two types of qubits in fault-tolerant quantum computing \cite{nielsen2000quantum}, one for the storage and the computation of the logical states and the other for the runtime detection and the correction of the error syndromes. For trapped ions, one of the leading platforms for quantum computing, these two types of qubits need to be spectrally separated to avoid crosstalk on one type due to the scattered photons during the measurement of the other \cite{tan2015multi,inlek2017multispecies,negnevitsky2018repeated,bruzewicz2019dual,Wan875}. Besides, such ancilla ions are also required to provide sympathetic cooling for the computational ions and help stabilize the trapped ion system as the qubit number increases \cite{PhysRevA.65.040304,PhysRevA.68.042302,PhysRevA.79.050305}. Furthermore, ancilla ions also play a pivotal role in the photonic quantum network scheme for scaling up the ion trap quantum computers by continually generating ion-photon entanglement \cite{10.5555/2011617.2011618, duan2010colloquium,PhysRevA.89.022317}.

Previously it was assumed that such frequency-separated dual-type ion qubits have to be implemented in hybrid systems of two ion species, which has attracted large experimental efforts with remarkable progress \cite{tan2015multi,inlek2017multispecies,negnevitsky2018repeated,bruzewicz2019dual,Wan875}. For hybrid systems, apart from the experimental complexity to trap and cool two ion species and the lower mixed-species gate ﬁdelity than the same-species case, it is also challenging to control the fraction and the positioning of each qubit type in many-ion crystals. Moreover, the mass mismatch between the ion species makes it very difficult to realize sympathetic cooling and high-fidelity gates with the transverse phonon modes \cite{sosnova2021character}, a choice that is necessitated for more scalable quantum gates in larger ion crystals \cite{Zhu2006PRL-transverse-mode,PhysRevA.89.022317,Lin_2009}.

\begin{figure*}[!tbp]
	\centering
	\includegraphics[width=0.8\linewidth]{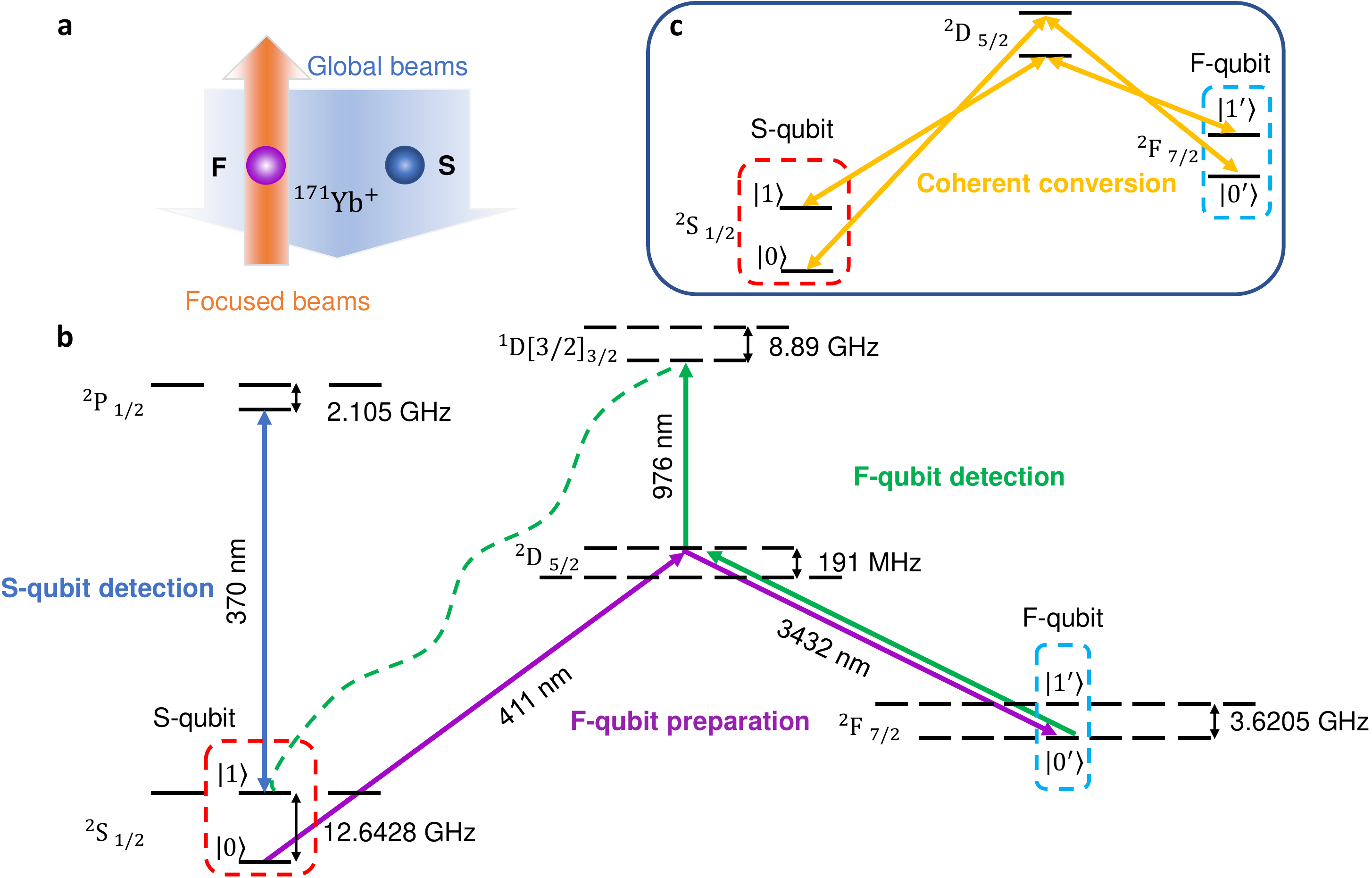}
	\caption { \textbf{Schematic of the experiment.} \textbf{a}, We use two trapped ${}^{171}\mathrm{Yb}^+$ ions to demonstrate the dual-type qubits. A focused narrow-band laser beam is used to convert the qubit types for a selected ion. Broad global laser beams can then operate the two types of qubits separately without affecting each other. \textbf{b}, Relevant energy levels of ${}^{171}\mathrm{Yb}^+$ and the schematic paths for the initialization and the detection of the dual-type qubits. The S-qubit is encoded in the commonly used clock states $|0\rangle$ and $|1\rangle$ of the $S_{1/2}$ levels, with routine qubit operations. For example, the detection process uses $370\,$nm laser (blue) together with $935\,$nm laser for repumping from the $D_{3/2}$ states (not shown). The F-qubit is encoded in another pair of clock states $|0^\prime\rangle$ and $|1^\prime\rangle$ of the $F_{7/2}$ levels. The F-qubit is initialized by first preparing the ion in $|0\rangle$ and is then transferred to $|0^\prime\rangle$ by successive $\pi$ pulses of $411\,$nm and $3432\,$nm laser (purple). To detect the F-qubit, we apply continuous $3432\,$nm and $976\,$nm laser (green) to optically pump the population in $|0^\prime\rangle$ back to $S_{1/2}$, which is a bright state under $370\,$nm laser, while $|1^\prime\rangle$ remaining a dark state. \textbf{c}, Coherent conversion between the S-qubit and the F-qubit. The two transfer paths $|0\rangle\leftrightarrow|0^\prime\rangle$ and $|1\rangle\leftrightarrow|1^\prime\rangle$ are traversed simultaneously by turning on suitable sidebands for the hyperfine splitting in the $411\,$nm and the $3432$nm $\pi$ pulses.}\label{Fig1}
\end{figure*}

In this paper, we experimentally realize dual-type qubits that are coherently convertible to each other with the same species of ${}^{171}\mathrm{Yb}^+$ ions. Coherent conversion between different qubit types allows us to dynamically tune the fraction and the positioning of each qubit type on demand in many-ion crystals during the computation, which is highly desirable for efficient sympathetic cooling \cite{lin2016sympathetic} and quantum error correction \cite{nielsen2000quantum,campbell2017roads} in large-scale systems. In addition, the capability of fast and high-fidelity qubit type conversion indicates that entangling gates between different qubit types can be performed in exactly the same way as the gates within the same qubit types, hence eliminating a challenging requirement of mixed-species high-fidelity gates. Both types of qubits in our experiment are realized with clock states of ${}^{171}\mathrm{Yb}^+$ ions, in the S and F manifolds, respectively, which have long coherence time and almost no relaxation. The coherent conversion is achieved with dual-tone narrow-band laser beams at wavelengths of $411\,$nm and $3432\,$nm, respectively, with a fidelity about $99.7\%$ already in the first experiment. We then demonstrate that, during the operations on one qubit type such as cooling, gates and detection, the coherence of the other qubit type is well preserved with the crosstalk error rate below $0.03\%$ per operation. Note that the transition to the meta-stable D or F levels have been used before in ion trap experiments for qubit detection through electron shelving \cite{leibfried2003quantum,edmunds2020scalable,christensen2020high} and for temporary protection of the optical qubits under detection \cite{barreiro2011open,Monz1068}. The contribution here is that we achieve the first fast coherent conversion between dual types of qubits both carried by robust clock states with long coherence time and almost no relaxation. This allows us to realize the required whole set of protected operations through dual types of qubits with the below-threshold crosstalk error rates, including the first demonstration of the sympathetic cooling using the same ion species with negligible crosstalk errors.

To demonstrate the protection by the use of dual types of qubits, it suffices to consider two ions. Our experimental setup consists of two ${}^{171}\mathrm{Yb}^+$ ions, as shown schematically in Fig.~\ref{Fig1}. Each ion can be in one of the two qubit types, encoded either in the clock states $|0\rangle\equiv |F=0,m_F=0\rangle$ and $|1\rangle\equiv |F=1,m_F=0\rangle$ of the $\mathrm{S}_{1/2}$ levels (S-qubit), or $|0^\prime\rangle\equiv |F=3,m_F=0\rangle$ and $|1^\prime\rangle\equiv |F=4,m_F=0\rangle$ of the metastable $\mathrm{F}_{7/2}$ levels (F-qubit). The S-qubit can be manipulated by routine laser and microwave operations \cite{PhysRevA.76.052314} such as optical pumping, qubit state detection, Doppler cooling and single-qubit gates. For a single S-qubit, we achieve $98.3\%$ fidelity for detection using $370\,$nm laser and $99.98\%$ fidelity for single-qubit rotations via $12.6\,$GHz microwave, which suffice for our experiment. If higher detection fidelity is needed, we can apply the electron shelving technique \cite{leibfried2003quantum,edmunds2020scalable,christensen2020high} and have gotten a detection fidelity above $99.9\%$ in our experiment. More details can be found in Methods.

To prepare an F-qubit, we first initialize the ion in $|0\rangle$ through optical pumping and then transfer its population to $|0^\prime\rangle$ via the intermediate state of $\mathrm{D}_{5/2}$ using a $411\,$nm $\pi$ pulse followed by another $3432\,$nm $\pi$ pulse (from a house-made narrow-band laser, see Methods). The $411\,$nm laser is focused to a beam waist radius of about $4\,\mu$m and supports selective control of one ion with small crosstalk on the other ion at a distance of about $14\,\mu$m. Because of the finite laser linewidth, the population transfer fidelity does not reach unity. Therefore we add a verification step to check if the ion remains in $\mathrm{S}_{1/2}$ or $\mathrm{D}_{5/2}$ levels, and discard these unsuccessful events. The F-qubit can then be operated by $3.6\,$GHz microwave with a single-qubit gate fidelity of $99.99\%$ (see Methods). For its detection, we perform electron shelving by incoherently pumping the population in $|0^\prime\rangle$ back to the $S_{1/2}$ levels through continuous $3432\,$nm and $976\,$nm laser with $20\,\mu$s duration. In this way, $|0^\prime\rangle$ is mapped to a bright state under $370\,$nm detection laser while $|1^\prime\rangle$ remains a dark state, which gives us a detection fidelity of $99.86\%$ at $250\,\mu$s detection time. Higher detection ﬁdelity of 99.97\% is also demonstrated in our experiment using longer detection time (see Methods).

\begin{figure}[!tbp]
	\centering
	\includegraphics[width=\linewidth]{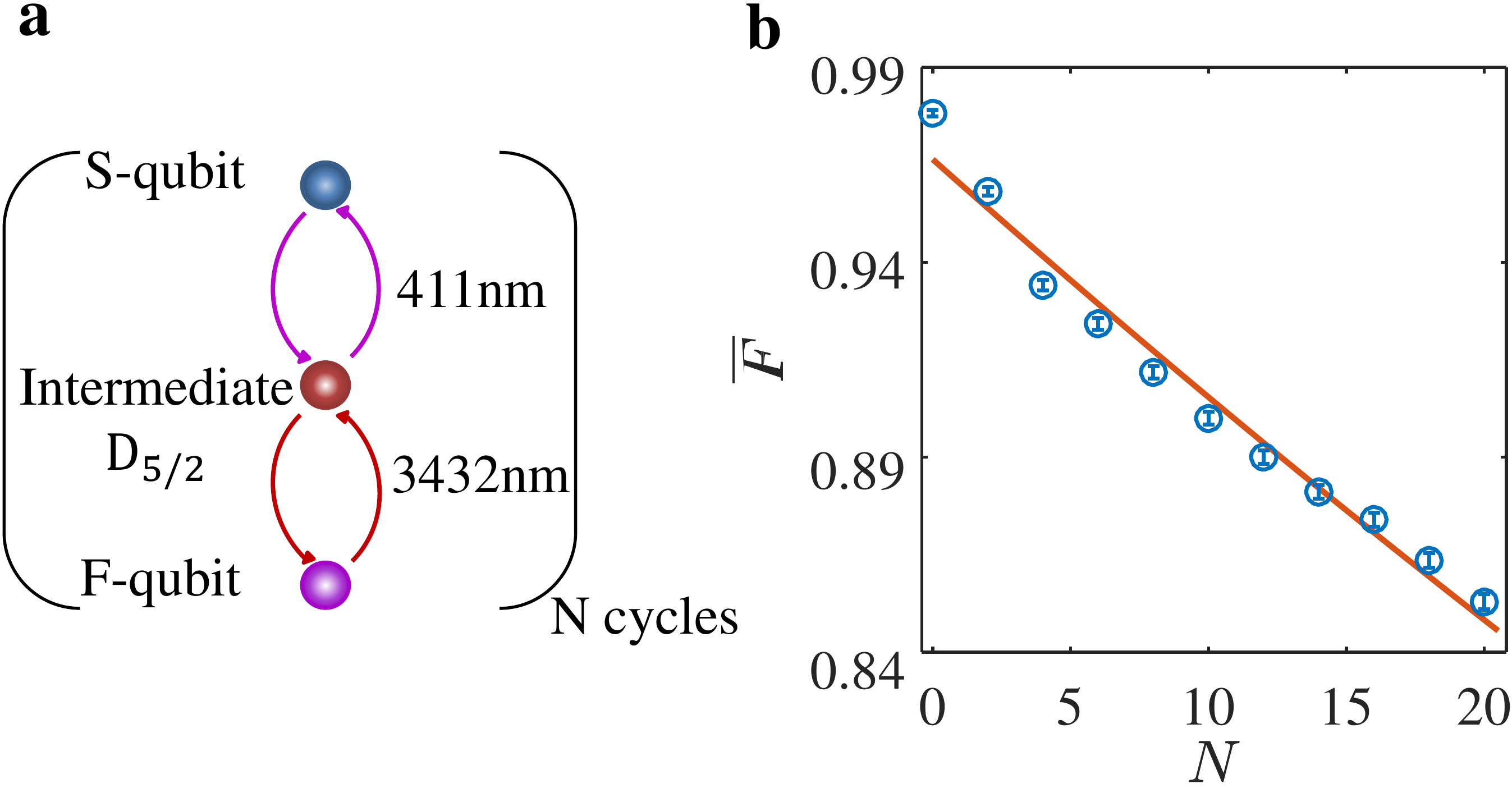}
	\caption {\textbf{Coherent conversion between the two qubit types.} \textbf{a}, We use the transfer paths in Fig.~\ref{Fig1}c to coherently convert an S-type qubit to the F-type and then back to the S-type, and repeat this cycle for $N$ times. To further increase the fidelity of the conversion, sideband cooling is applied to reduce phonon numbers (see Methods). \textbf{b}, We initialize the qubit in the mutually unbiased bases (MUBs) $|0\rangle$, $|1\rangle$, $|+\rangle$, $|-\rangle$, $|L\rangle$ and $|R\rangle$, and measure the average fidelity between the initial and the final states after $N$ rounds of S-F-S conversion. The average fidelity is fitted as $\overline{F}=F_0 (1-\epsilon_t)^{N}$ where $F_0\approx (96.6\pm0.4)\%$ characterizes the state preparation and measurement (SPAM) fidelity and $\epsilon_t\approx (0.65\pm0.03)\%$ for the round-trip transfer error. Error bars represent one standard deviation and are smaller than the data marker in the plot.}\label{Fig2}
\end{figure}

The two qubit types can be converted into each other coherently in less than one microsecond. As shown in Fig.~\ref{Fig1}\textbf{c}, an S-qubit can first be transferred to the $\mathrm{D}_{5/2}$ levels through a $411\,$nm $\pi$ pulse with suitable microwave sidebands for $|0\rangle$ and $|1\rangle$ simultaneously. Then another $3432\,$nm $\pi$ pulse, again with suitable microwave sidebands, finishes the conversion to the F-qubit. By reversing the order of the two $\pi$ pulses we can similarly achieve the conversion from an F-qubit back to the S-type. During this process, the phase noise in the laser beams appears as a global phase for the qubit and thus does not lead to decoherence.
In Fig.~\ref{Fig2} we use the fidelity between the qubit states before and after the coherent conversion to quantify its performance. Specifically, we initialize an S-qubit state $|\psi\rangle$ through a microwave pulse, perform $N$ rounds of S-F-S conversions, and measure the fidelity of the final state $\rho$ with the initial state $\langle\psi|\rho|\psi\rangle$ by mapping $|\psi\rangle$ back to $|1\rangle$ using another microwave pulse followed by detection in the $|0\rangle$/$|1\rangle$ basis.
By averaging the initial state $|\psi\rangle$ over a complete set of mutually unbiased bases (MUBs) \cite{WOOTTERS1989363} $|0\rangle$, $|1\rangle$, $|+\rangle\equiv(|0\rangle+|1\rangle)/\sqrt{2}$, $|-\rangle\equiv(|0\rangle-|1\rangle)/\sqrt{2}$, $|L\rangle\equiv(|0\rangle+i|1\rangle)/\sqrt{2}$ and $|R\rangle\equiv(|0\rangle-i|1\rangle)/\sqrt{2}$, we get the average fidelity $\overline{F}$ over the Bloch sphere \cite{Klappenecker2005} which is fitted as $\overline{F}=F_0 (1-\epsilon_t)^{N}$. We thus extract about $(3.4\pm0.4)\%$ state preparation and measurement (SPAM) error from $F_0$, and about $(0.65\pm0.03)\%$ coherent transfer error for each round of S-F-S conversion (or about $0.3\%$ error for one-way S-to-F or F-to-S conversions), with error bars representing one standard deviation.

\begin{figure}[!tbp]
	\centering
	\includegraphics[width=\linewidth]{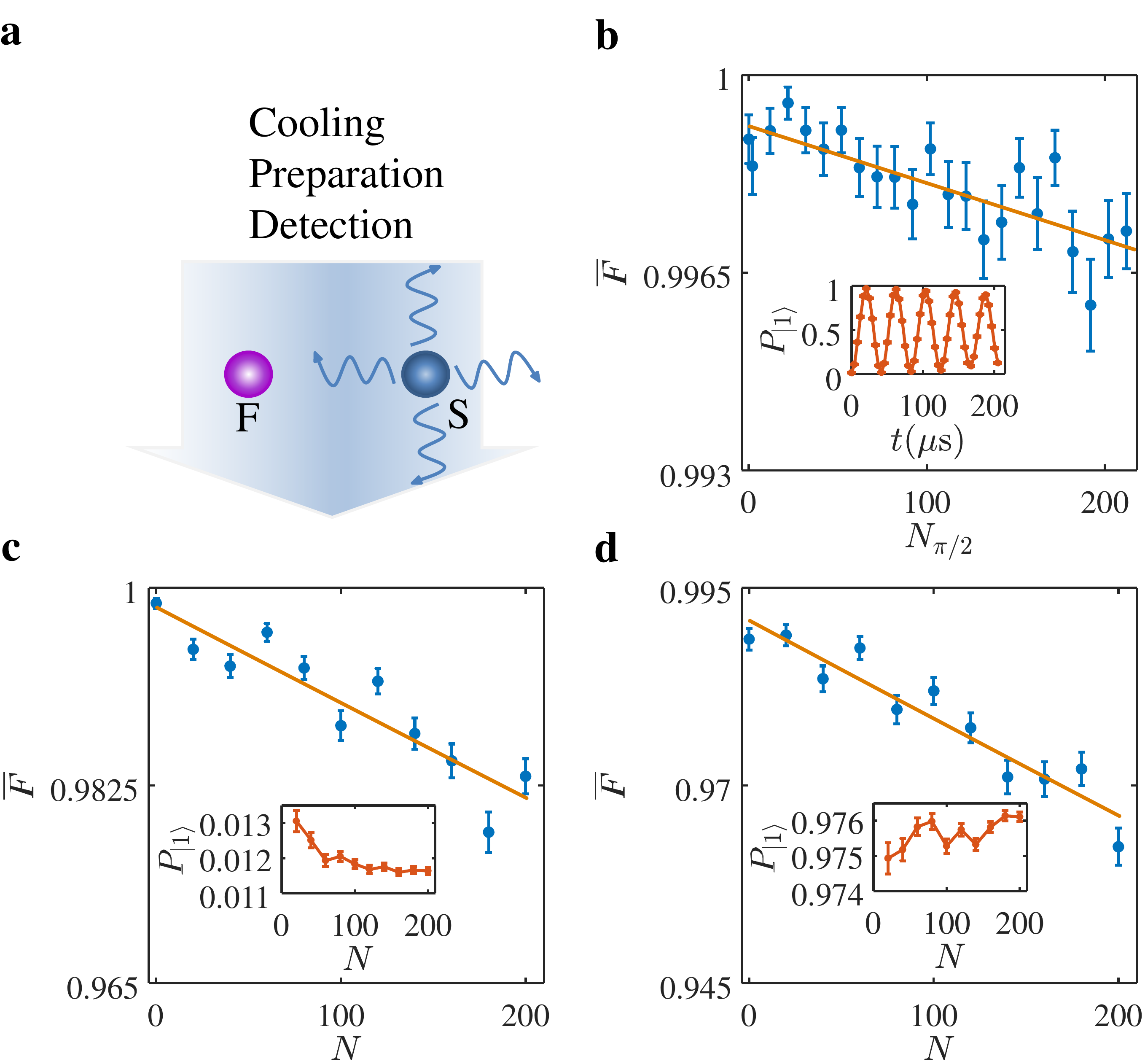}
	\caption {\textbf{Crosstalk error of S-qubit operations on the F-qubit.} \textbf{a}, We prepare an S-qubit and an F-qubit. The cooling, detection and gate operations on the S-qubit will not affect the F-qubit owing to the difference in their operating frequencies. \textbf{b}, We initialize the S-qubit in $|0\rangle$, drive the Raman transition between $|0\rangle$ and $|1\rangle$ using $355\,$nm laser, and measure the population in $|1\rangle$ versus the evolution time. As shown in the inset, we observe a Rabi oscillation and the decay mainly comes from the decoherence of the laser. The average fidelity $\overline{F}$ of the F-qubit in the MUBs drops slowly with the number of $\pi/2$ Rabi pulses $N_{\pi/2}$ and is fitted as $\overline{F}=F_0^r (1-\epsilon_r)^{N_{\pi/2}}$ where $F_0^r\approx (99.91\pm0.02)\%$ again represents the SPAM fidelity and $\epsilon_r\approx (0.0010\pm0.0002)\%$ is the crosstalk error per $\pi/2$ pulse. \textbf{c}, We apply Doppler cooling, optical pumping to $|0\rangle$, and detection on the S-qubit and repeat this cycle for $N$ times. The average fidelity of the F-qubit versus $N$ is fitted as $\overline{F}=F_0^0 (1-\epsilon_0)^{N}$ with $F_0^0\approx (99.8\pm0.1)\%$ and $\epsilon_0\approx (0.009\pm0.001)\%$. The measured probability of the S-qubit in $|1\rangle$ is shown in the inset, whose deviation from zero represents the SPAM error. \textbf{d}, We add a microwave $\pi$ pulse in the experimental sequence of (c) after optical pumping to prepare the S-qubit in $|1\rangle$. Here we fit $\overline{F}=F_0^1 (1-\epsilon_1)^{N}$ with $F_0^1\approx (99.1\pm0.2)\%$ and $\epsilon_1\approx (0.013\pm0.001)\%$. The deviation of the inset from one comes from the SPAM error. All error bars represent one standard deviation.} \label{Fig3}
\end{figure}

In Fig.~\ref{Fig3} we analyze the crosstalk error of the S-qubit operations on a nearby F-qubit. We begin with two S-qubit ions in $|0\rangle$ after Doppler cooling and optical pumping, and then initialize one ion into an F-qubit in the state $|\psi^\prime\rangle$. As shown in Fig.~\ref{Fig3}\textbf{b}, we apply co-propagating $355\,$nm Raman laser beams \cite{PhysRevLett.112.190502} to drive the resonant transition between $|0\rangle$ and $|1\rangle$ of the S-qubit, with the Rabi oscillation presented in the inset. The average fidelity $\overline{F}$ of the F-qubit over the MUBs decays slowly with the duration $t$ of the Raman laser, or measured in terms of the number of $\pi/2$ pulses $N_{\pi/2}\equiv t/\tau_{\pi/2}$ where $\tau_{\pi/2}$ is the required Raman laser duration to achieve a $\pi/2$ pulse. We fit the data by $\overline{F}=F_0^r (1-\epsilon_r)^{N_{\pi/2}}$ and extract a crosstalk error of $\epsilon_r\approx (0.0010\pm0.0002)\%$ on the F-qubit for a $\pi/2$ Raman pulse on the S-qubit. In Fig.~\ref{Fig3}\textbf{c}, after initializing the F-qubit, we turn on the Doppler cooling, optical pumping and state detection cycle for the S-qubit, and repeat this sequence for $N$ times after which we measure the F-qubit fidelity. As we can see from the inset, we have about $1\%$ SPAM error when preparing an S-qubit in $|0\rangle$ and detecting in $|1\rangle$, which mainly comes from the imperfect S-qubit detection. The error bar in the inset is reducing for larger $N$ because we are averaging over more experimental trials, but the mean value should stay constant; the weak decreasing tendency in the plot may be due to the slow parameter drifts in the experiment. In this case we fit the average F-qubit fidelity $\overline{F}=F_0^0 (1-\epsilon_0)^{N}$ with $\epsilon_0\approx (0.009\pm0.001)\%$ as the crosstalk error on the F-qubit when initializing and detecting a $|0\rangle$ state S-qubit. Similarly, in Fig.~\ref{Fig3}\textbf{d} we add a microwave $\pi$ pulse in the experimental sequence to initialize the S-qubit in $|1\rangle$ and we fit a crosstalk error of $\epsilon_1\approx(0.013\pm 0.001)\%$.

\begin{figure}[!tbp]
	\centering
	\includegraphics[width=0.9\linewidth]{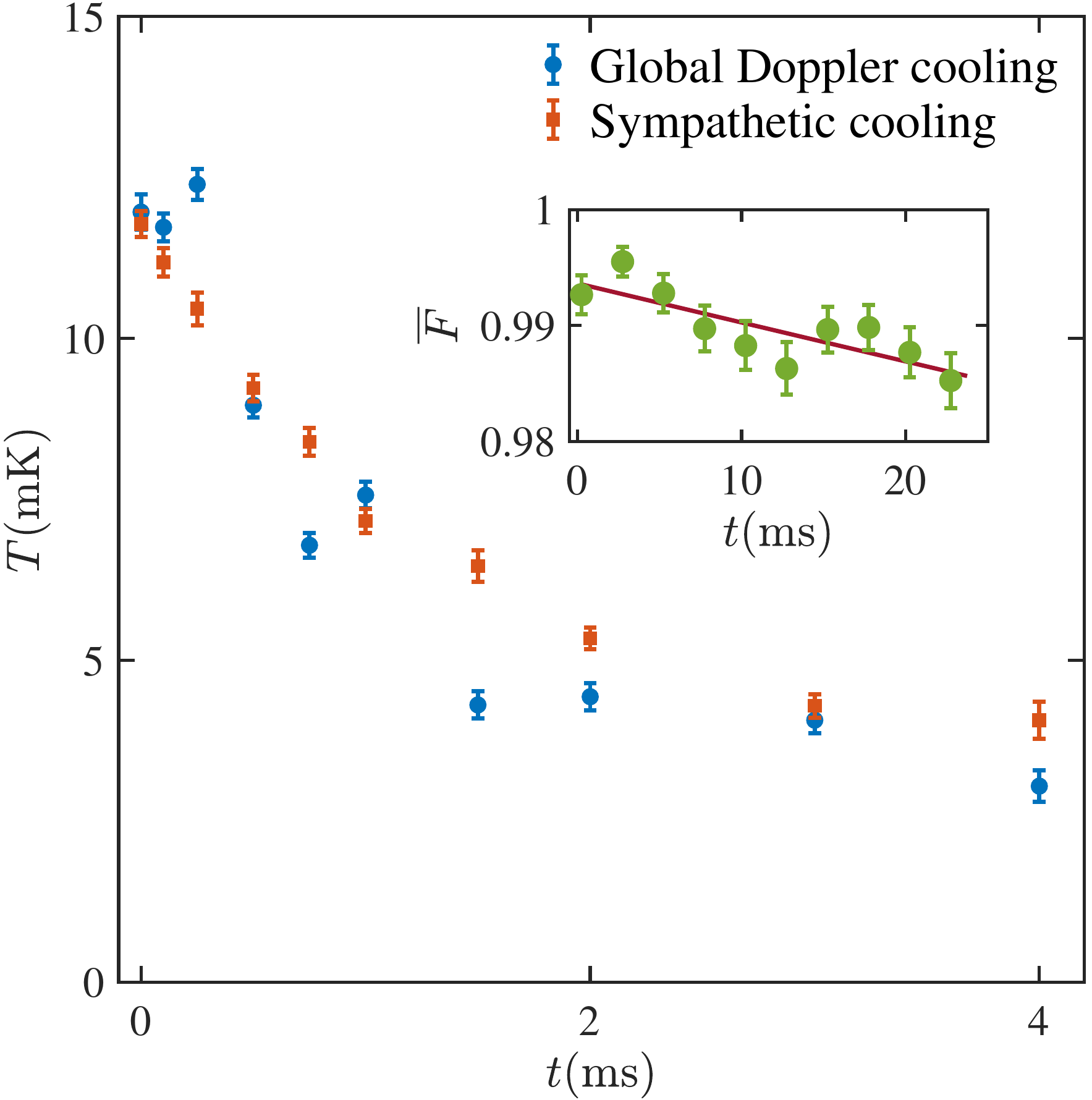}
	\caption {\textbf{Sympathetic cooling and crosstalk error.} We prepare an S-type qubit and an F-type qubit, and apply Doppler cooling on the S-qubit using $370\,$nm laser with $10\,$MHz red detuning. Starting from a high temperature, we observe a decay in the measured average temperature versus the cooling time (red square). In comparison, blue dots are the measured average temperature under global Doppler cooling when both ions are in the S-type. The cooling rates and the final temperatures are comparable in the two cases. The inset shows the average fidelity of the F-qubit over the MUBs when the Doppler cooling laser is on. Fitting with an exponential function $\overline{F}=F_0^c e^{-t/T_c}$, we get a decoherence time $T_c=(2.9\pm0.9)\,$s, which corresponds to about $0.03\%$ crosstalk error during a sympathetic cooling operation of a typical $1\,$ms duration. Error bars represent one standard deviation.}\label{Fig4}
\end{figure}

The Doppler cooling stage in the sequence above turns out to be important for our experiment because otherwise the spatial motion of the ions will be heated during the evolution time of tens of milliseconds, which will hinder the accuracy of the subsequent laser manipulations. In Fig.~\ref{Fig4} we explicitly study this sympathetic cooling dynamics. Before initializing the F-qubit when both ions are in the S-type, first we heat them by blue-detuned $370\,$nm laser. Then we prepare an F-qubit in $|0^\prime\rangle$ and turn on the Doppler cooling beam for the S-qubit for various duration. Finally we repump the F-qubit back to the S-type $|0\rangle$ to measure the temperature of the two-ion crystal through the carrier Rabi oscillation of the sideband-resolved $411\,$nm transition (see Methods). In comparison, we also apply the heating-cooling sequence without preparing the F-qubit such that both ions can be cooled by the Doppler cooling beam. As we can see, the cooling dynamics for the sympathetic cooling and the global Doppler cooling cases are similar, which proves the efficiency of the sympathetic cooling. In the inset, we further measure the average F-qubit fidelity $\overline{F}$ (similarly over the MUBs) versus the sympathetic cooling time $t$ and fit $\overline{F}=F_0^c e^{-t/T_c}$ to get a decoherence time $T_c=(2.9\pm0.9)\,$s. This corresponds to about $0.03\%$ crosstalk error for a sympathetic cooling operation of a typical $1\,$ms duration.
Note that in our experiment, the $411\,$nm laser is perpendicular to the two-ion chain, so it probes the temperature of transverse phonon modes.
It has been shown that sympathetic cooling of the transverse modes is inefficient in hybrid systems using two ion species due to localization of the modes caused by the mass mismatch \cite{sosnova2021character}, but our dual-type qubits within the same ion species allow efficient cooling for these modes, which can be used for more robust entangling gates in larger ion crystals and thus advantageous for scalable quantum computing \cite{Zhu2006PRL-transverse-mode,PhysRevA.89.022317,Lin_2009}.

In summary, we have experimentally demonstrated dual-type qubits within the same ion species. The two qubit types can be coherently converted into each other using microsecond narrow-band laser pulses, with about $99.7\%$ transfer fidelity each way. The fidelity is currently limited by the imperfect population transfer due to the technical noise of the laser power and frequency fluctuation and can be improved by better frequency locking of the laser or using more robust composite pulses \cite{PhysRevA.70.052318}. Crosstalk errors from sympathetic cooling, optical pumping, detection and single-qubit gates on the S-qubit are measured to be all below $0.03\%$ for the F-qubit. The demonstrated below-threshold crosstalk errors between the dual types of qubits, together with their fast high-fidelity coherent conversion, opens up prospects of wide applications in large-scale quantum computing and quantum networking.

\bigskip

\textbf{Data Availability:} The data that support the findings of this study are available
from the authors upon request.

\textbf{Acknowledgements:} This work was supported by the Beijing Academy of Quantum Information Sciences, the National key Research and Development Program of China, Frontier Science Center for Quantum Information of the Ministry of Education of China, and Tsinghua University Initiative Scientific Research Program. Y.-K. W. acknowledges support from Shuimu Tsinghua Scholar Program and International Postdoctoral Exchange Fellowship Program (Talent-Introduction Program).

\textbf{Competing interests:} The authors declare that there are no competing interests.

\textbf{Author Information:} Correspondence and requests for materials should be addressed to L.M.D.
(lmduan@tsinghua.edu.cn).

\textbf{Author Contributions:} L.M.D. proposed and supervised the experiment. H.X.Y., J.Y.M., Y.W., M.M.C., W.X.G., Y.Y.H., F.L., Y.K.W., Z.C.Z. carried out the experiment. H.X.Y., J.Y.M., Y.K.W., and L.M.D. wrote the manuscript.

\makeatletter
\renewcommand\@biblabel[1]{#1.}
\makeatother

\renewcommand{\figurename}{Extended Data Fig.}
\setcounter{figure}{0}
\makeatletter
\apptocmd{\thebibliography}{\global\c@NAT@ctr 30\relax}{}{}
\makeatother

\section{Methods}
\subsection{Experimental setup}
In this experiment, we use a single $370\,$nm laser beam which drives transitions between $\mathrm{S}_{1/2}$ and $\mathrm{P}_{1/2}$ of the ${}^{171}\mathrm{Yb}^+$ ions for Doppler cooling, optical pumping and detection of the S-qubit \cite{PhysRevA.76.052314}. To switch the role of the laser beam, we use an acousto-optical modulator (AOM) controlled by a home-made direct digital synthesizer (DDS) to quickly change the carrier frequency and we turn on different electro-optical modulators (EOMs) to generate the desired microwave sidebands. For Doppler cooling, the laser is set to be about $10\,$MHz red detuned from the $|\mathrm{S}_{1/2},F=1\rangle\leftrightarrow|\mathrm{P}_{1/2},F=0\rangle$ transition with a $14.7\,$GHz sideband for the $|\mathrm{S}_{1/2},F=0\rangle\leftrightarrow|\mathrm{P}_{1/2},F=1\rangle$ transition. For optical pumping into $|0\rangle$, the laser is set to be resonant with the $|\mathrm{S}_{1/2},F=1\rangle\leftrightarrow|\mathrm{P}_{1/2},F=0\rangle$ transition with a $2.1\,$GHz sideband for the $|\mathrm{S}_{1/2},F=1\rangle\leftrightarrow|\mathrm{P}_{1/2},F=1\rangle$ transition.
The detection beam is resonant to the $|\mathrm{S}_{1/2},F=1\rangle\leftrightarrow|\mathrm{P}_{1/2},F=0\rangle$ transition with no microwave sideband. The optical power for this laser beam is about $10\,\mu$W with a beam waist diameter (where the intensity drops to $1/e^2$) of about $60\,\mu$m. The small population leaked to the $\mathrm{D}_{3/2}$ levels is pumped back to $\mathrm{S}_{1/2}$ by $935\,$nm laser, with an optical power around $1\,$mW and a beam waist diameter around $100\,\mu$m.

\begin{figure*}[htbp]
	\centering
	\includegraphics[width=0.8\linewidth]{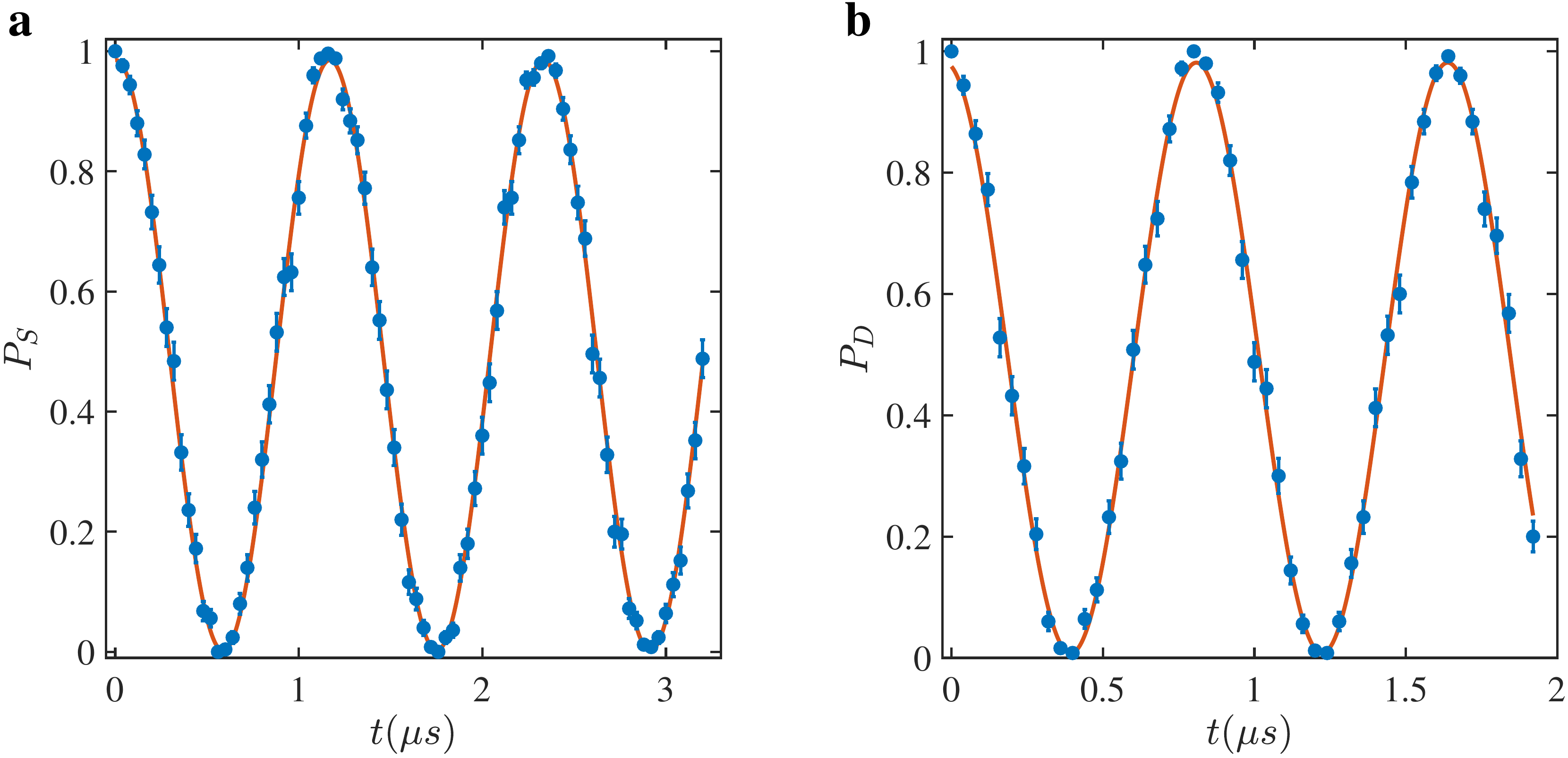}
	\caption {Carrier Rabi oscillation of \textbf{a}, the $411\,$nm laser and \textbf{b}, $3432\,$nm laser. The $411\,$nm laser has an optical power of about $0.8\,$mW and a beam waist diameter of about $8\,\mu$m, which generates a Rabi frequency of about $2\pi\times859.4\,$kHz. The $3432\,$nm laser has an optical power of about $0.5\,$mW and a beam waist diameter of about $73\,\mu$m, which gives a Rabi frequency of about $2\pi\times 1.2\,$MHz.}
	\label{FigS1}
\end{figure*}

The conversion between the S-qubit and the F-qubit is accomplished via $411\,$nm and $3432\,$nm laser to bridge $\mathrm{S}_{1/2}\leftrightarrow \mathrm{D}_{5/2}$ and $\mathrm{D}_{5/2}\leftrightarrow \mathrm{F}_{7/2}$ transitions, respectively. The frequency of the $411\,$nm laser is stabilized by a cavity, and its decoherence time is measured to be around $230\,\mu$s. It has an optical power of tens to thousands of $\mu$W and a beam waist diameter of about $8\,\mu$m, which generate a Rabi frequency of hundreds of kHz. This focused beam allows selective control of one ion with small crosstalk on the other ion.
The house-made $3432\,$nm laser, with optical power up to $1.1\,$W, is produced by difference-frequency generation using $30\,$W $1064\,$nm laser and $30\,$W $1542\,$nm laser in a $\mathrm{LiNbO}_3$ crystal \cite{Taylor1998Combined}. With the frequencies of the $1064\,$nm and the $1542\,$nm laser stabilized by the same cavity, a coherence time of about $20\,\mu$s for the $3432\,$nm laser is achieved.
In this experiment, we only use an optical power of several mW for the $3432\,$nm laser, and get a Rabi frequency of a few MHz. The carrier Rabi oscillations under typical parameters for these two laser beams are presented in Extended Data Fig.~\ref{FigS1}.

To maintain the coherence during the qubit type conversion, we drive the two transition paths for the two basis states of the qubit simultaneously. This is achieved by using the two first-order sideband frequency components generated by an EOM. We tune the carrier frequency of the $411\,$nm ($3432\,$nm) laser to the central frequency of the $\mathrm{S}_{1/2}\leftrightarrow \mathrm{D}_{5/2}$ ($\mathrm{D}_{5/2}\leftrightarrow \mathrm{F}_{7/2}$) transitions (see Fig.~1c for the two paths), and set the driving frequency on the EOM to be half of the frequency difference between the two paths, that is, $6.42\,$GHz ($1.91\,$GHz), respectively.

To repump the $\mathrm{D}_{5/2}$ states back to $\mathrm{S}_{1/2}$ and to incoherently transfer the population from $|0^\prime\rangle$ (together with a continuous $3432\,$nm laser), we use a $1.5\,$mW $976 \,$nm laser beam to drive the $\mathrm{D}_{5/2}\leftrightarrow [3/2]_{3/2}$ transition, which can depump the $\mathrm{D}_{5/2}$ states in about $10\,\mu$s.

We use a blade trap to hold the ions. To measure the crosstalk between two ions, we use trap frequencies $\omega_x=2\pi\times 3.15\,$MHz, $\omega_y=2\pi\times 2.97\,$MHz and $\omega_z=2\pi\times 120\,$kHz such that the ion separation is about $14\,\mu$m. For the coherent conversion of qubit types, similar trap parameters and Doppler cooling lead to a round-trip transfer error of about $(1.98\pm0.05)\%$, which we believe is influenced by the large phonon number. Therefore we slightly raise the transverse trap frequencies to $\omega_x=2\pi\times 3.63\,$MHz and $\omega_y=2\pi\times 3.48\,$MHz, and apply preliminary sideband cooling using $411\,$nm laser to reduce the average phonon number to about 3. Then we achieve about $(0.65\pm 0.03)\%$ round-trip transfer error as reported in the main text using $0.54\,\mu$s $411\,$nm $\pi$ pulses and $0.39\,\mu$s $3432\,$nm $\pi$ pulses.

\subsection{Detection through Electron shelving}
In our experiment, the F-qubit is detected by electron shelving as shown in the main text with a detection fidelity of $(99.86\pm0.03)\%$ at $0.25$ ms detection time, which can be further improved to $(99.97\pm0.004\%)$ at $2.5$ ms detection time. The S-qubit can be detected by the routine $370\,$nm laser with a detection fidelity of $(98.3\pm0.2)\%$, or by the electron shelving method \cite{edmunds2020scalable} with a detection fidelity of $(99.91\pm0.007)\%$.

Similar to the preparation of an F-qubit, we can transfer the population in $|0\rangle$ of the S-qubit to the $|\mathrm{F}_{7/2},F=3,m_F=0\rangle$ state via $411\,$nm and $3432\,$nm $\pi$ pulses. However, this population transfer is not perfect, so we further transfer the population in $|0\rangle$ to several Zeeman levels of the $|\mathrm{F}_{7/2},F=4,m_F=0,\pm 1\rangle$ states sequentially using $3.62\,$GHz microwaves. Then we achieve a detection fidelity of around $(99.913\pm0.007)\%$ for the S-qubit within a detection time of about $2.5$ ms.

\subsection{Microwave single-qubit gates}
\begin{figure*}[htbp]
	\centering
	\includegraphics[width=0.8\linewidth]{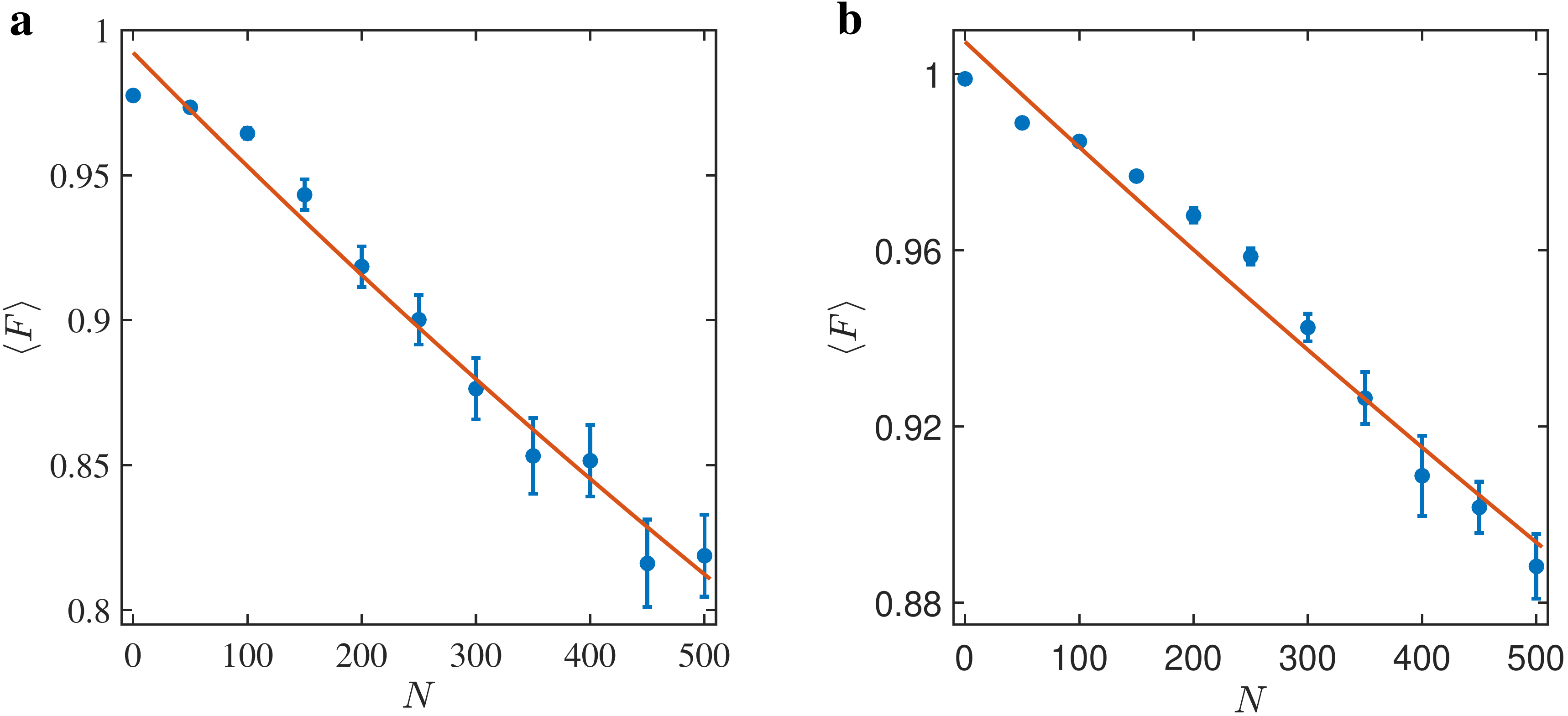}
	\caption {Randomized benchmarking of the microwave-driven single-qubit gates for \textbf{a}, the S-qubit and \textbf{b}, the F-qubit. The average gate fidelity is $(99.98\pm0.04)\%$ for the S-qubit and $(99.99\pm0.04)\%$ for the F-qubit.}
	\label{FigS2}
\end{figure*}

We use $12.6428\,$GHz microwave fields to implement single-qubit gates for the S-qubit, and $3.6205\,$GHz microwave fields for the F-qubit. We characterize the gate fidelity via the standard randomized benchmarking method \cite{magesan2011scalable}. The results are shown in Extended Data Fig.~\ref{FigS2} with the measured average gate fidelity of $(99.98\pm0.04)\%$ for the S-qubit and $(99.99\pm0.04)\%$ for the F-qubit. Therefore we can ignore the error in the microwave operations in the initialization and the measurement stages compared with the other SPAM errors.

\subsection{Fit average temperature}
We use the decay of the carrier Rabi oscillation amplitude of the $411\,$nm laser to investigate the motional information of the two-ion chain \cite{feng2020efficient}. The propagation direction of the $411\,$nm laser is perpendicular to the axial direction and is at an angle of $45^\circ$ to both the $x$ and the $y$ axes, thus it probes the four transverse modes (the stretch mode and the common mode in each direction).
Here we define an effective temperature $T$ under the assumption that all the motional modes share the same temperature. The average phonon number for mode $i$ with frequency $\omega_i$ is then
$\overline{n}_i = 1/(e^{\hbar\omega_i/k_BT}-1)$.
The probability that the four modes have phonon number $\{n_i\} = \{n_1,n_2,n_3,n_4\}$ is thus $P_{\{n_i\}}=\Pi_{i=1}^{4}\overline{n}_i^{n_i}/(\overline{n}_i+1)^{n_i+1}$.

The carrier Rabi frequency under $\{n_i\}$ is \cite{leibfried2003quantum} $\Omega_{\{n_i\}} = \Omega_0\Pi_{i=1}^4 e^{-\eta_i^2/2}L_{n_i}(\eta_i^2) \approx\Omega_0 [1-\sum_{i=1}^4(n_i+1/2)\eta_i^2]$
where we have dropped the higher order terms of $n\eta^2$, since in our case $\eta\approx0.024$.

\begin{figure*}[htbp]
	\centering
	\includegraphics[width=0.5\linewidth]{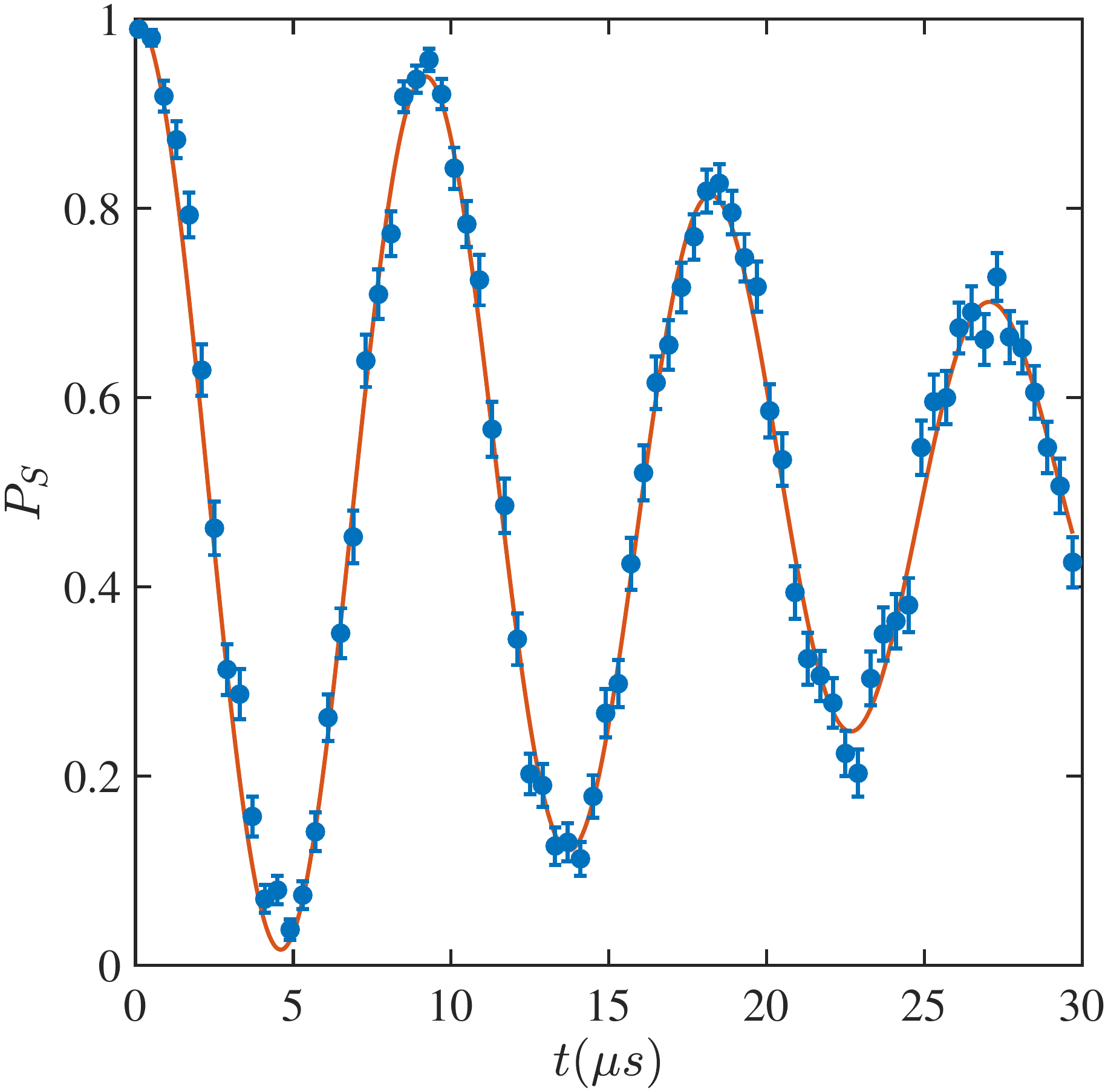}
	\caption {Carrier Rabi oscillation of $411\,$nm laser after $500\,\mu$s sympathetic cooling. The fitted effective temperature is about $(9.2\pm 0.2)\,$mK.}
	\label{FigS3}
\end{figure*}

Now if we initialize the ion in the $\mathrm{S}_{1/2}$ state, its probability to remain in $\mathrm{S}_{1/2}$ is $P_S(t) = \sum_{\{n_i\}} P_{\{n_i\}}(1+\mathrm{cos}\Omega_{\{n_i\}}t)/2
=(1+\mathrm{Re}[f(t)])/2$ where $f(t) \equiv e^{i\Omega_0 t (1-\sum_i \eta_i^2 / 2)} \prod_i 1 / (\overline{n}_i+1-\overline{n}_i e^{-i \eta_i^2 \Omega_0 t })$.
We use it to fit the two parameters $\Omega_0$ and $T$ from the carrier Rabi oscillation. An example is shown in Extended Data Fig.~\ref{FigS3}.

\end{document}